\documentclass[12pt,preprint]{aastex}

\def\La{$Ly-\alpha$\,}

\newcommand\kms{km~s$^{-1}$}

\newcommand\grb{$\gamma$-ray burst}
\newcommand\grbs{$\gamma$-ray bursts}

\begin{document}

\title{Spectropolarimetry of GRB 021004 - Evidence for High Velocity \La\
Absorptions
\footnote{Based on observations collected at the
European Southern Observatory, Chile (ESO Progr. No. 68.D-0571(A)).}
}

\author{Lifan Wang$^{1}$, Dietrich Baade$^{2}$,
Peter H\"oflich$^{3}$, J. Craig Wheeler$^3$}

\affil{$^1$Lawrence Berkeley National Laboratory 50-232\\
    1 Cyclotron Rd, CA 94720}

\affil{$^2$European Southern Observatory\\
    Karl-Schwarzschild-Strasse 2\\
     D-85748 Garching, Germany}

\affil{$^3$Department of Astronomy and McDonald Observatory\\
          The University of Texas at Austin\\}

\begin{abstract}   
We present spectropolarimetry observations of GRB~021004 obtained at
the ESO VLT. We detect a remarkable increase of the degree of
polarization blueward of the host galaxy rest frame \La\ to
about 5-10\% that corresponds to a decrease in the continuum
flux below the extrapolation of a power law. An
increase of the degree of polarization is also recorded in some
narrow lines, but of lower statistical significance. 
The broad polarization feature blueward of the \La\ absorption line at 
4040 \AA\ is at least partially produced by hydrogen-rich material 
with velocities approaching 40,000 \kms\ located beyond the front of 
the afterglow shock, rather than due entirely to intergalactic \La\ 
absorption systems on the line of sight to the \grb. The presence of 
the broad \La\ absorption provides evidence that 
the \grb\ occurred inside the ejecta from a hydrogen-rich supernova 
that exploded before the \grb.

\end{abstract}


\section{Introduction}

Polarization measurements have been reported for a few \grbs\ such as
GRB 990510 (Covino et al. 1999; Wijers et al. 1999), GRB 990712
(Rol et al., 2000), GRB 011211 (Covino, et al. 2002a) 
GRB 020813 (Barth et al. 2002; Covino et al.
2002b), and GRB 021004 (Covino et al. 2002c, 2002d, Rol et al. 2002).
A highly significant variation in the polarization level is observed
for GRB 020813 with constant polarization position angle.

Several models for \grb\ polarization have been proposed. Gruzinov \&
Waxman (1999) argued that if the magnetic field is globally random but
with a large number of patches where the magnetic field is locally
coherent, polarization of up to $\sim 10\%$ is expected,
especially at early times. Ghisellini \& Lazzati (1999) and Sari (1999)
considered a geometrical configuration in which a collimated fireball is
observed slightly off-axis, and found that this break of symmetry results
in  significant polarization.

  GRB~021004 was discovered
by HETE II satellite at 12:06 UT on 2002 October 4
\citep{shirasaki02}.  Prompt discovery of the optical transient by
\citep{fox02a} allowed rapid response of spectroscopic observations.
\citep{fox02b} obtained the first spectra and identified two intervening
systems at z = 1.38 and 1.60 from \ion{Mg}{1} and \ion{Mg}{2} absorption.
\citet{erac02} confirmed those features, noted several \ion{Fe}{2}
absorptions at these redshifts, and pointed out four absorption lines at
$\sim$ 4633, 4664, 5109, and 5152 \AA\ (see also Sahu et al. 2002;
Castander et al. 2002).  These were identified by \citep{chorfil02} as
\ion{C}{4} and \ion{Si}{4} features at red shift of $\sim 2.3$.  Chornock
\& Filippenko also identified Ly$\alpha$ emission at z = 2.323 and
absorption components and perhaps Ly$\beta$ at similar red shifts.
\citet{salamanca02} identified four absorption components of \ion{C}{4} at
z = 2.295, 2.298, 2.230 and 2.237, noting that the total spread is about
3000 \kms .  The existence of various components in the lines shows  that
this \grb\ is associated with a star forming region and, very likely, is
associated with the explosion of a massive star (Schaefer et al. 2002,
 \citet{mirabal02}, \citet{savaglio02}). 
The lines could be formed in the wind
of the progenitor, which could be an O star, or a Wolf-Rayet (WR)
star which both
show typical expansion velocities of the order of 1,000 to 3,000 \kms .
Alternatively, \citet{salamanca02} suggested that the lines could be 
produced in a supernova shell that preceeded the explosion by several months.
This velocity dispersion would be difficult to interpret as due to a single
cluster of galaxies. For a more detailed
discussion see Schaefer et al. (2002).

Here we report polarization  spectra obtained with the ESO-VLT about 
5 days after the \grb\ which sheds new light on the origin and 
relation between massive stars and \grbs.

\section{Observations and Data Reductions}

The observations were made on 2002 Oct. 5 using the
Very Large Telescopes (UT3) of the European Southern Observatory.
Four exposures of 30 minutes each with the Wollaston Prism at position
angles at 0, 45, 22.5, and 65.5 were obtained using the ESO grism
GRIS$_300$V at UT Oct 5.202, 5.225, 5.248, and 5.269. No order sorting
filter was applied.  We used Melipal (= UT3) with FORS1 \citep{Wang01el:2002}.
FORS1 employs a TK2048EB4-1 2048$\times$2048 backside-illuminated
thinned CCD. The cooling of the CCD is performed by a standard
ESO bath cryostat. This system uses a FIERA controller for CCD
readout. The grism GRIS-300V (ESO number 10) was used for all
observations.  The range 360 to 850 nm was covered.

The data reduction package follows the descriptions given in the
ESO user manual. The Stokes parameters Q and U are converted to the degree
of polarization using the formula given in Wang et al (1996) which corrects
observational biases due to statistical noise. Systematic errors are
expected to be small ($<<\ 0.1\%$).

The flux spectrum and the Stokes $Q$ and $U$ parameters are shown in
Figure 1, where the data were binned to 15\AA\ to eliminate the
effect of correlations of the data values due to finite spectral
resolution.

\begin{figure}
\figurenum{1}
\epsscale{0.8}
\plotone{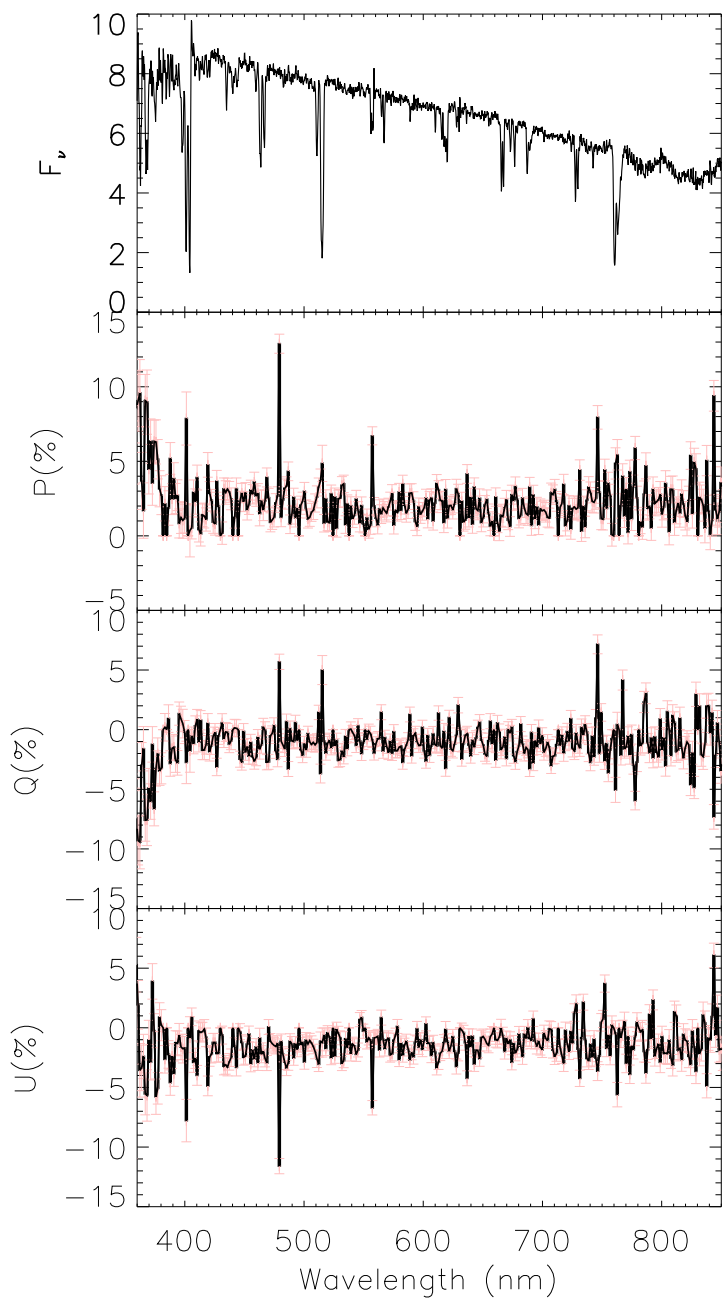} 
\caption{The flux spectrum (upper panel), polarization (second panel),
Stokes Q (third panel) and U (lower panel). 
The data are binned to 15 \AA\ which is slightly larger
than the instrumental resolution to reduce the effect of correlated
errors of neighboring pixel values. The errors are purely statistical, with
no account of systematics.
}
\end{figure} 

\section{Results from Observations}

\subsection{The Continuum Polarization}
 
Broad band polarization of GRB 021004 obtained at the ESO VLT-UT3
on October 5.080 to 5.149 was reported by Covino et al (2002c)
where it was found that the $V$ band polarization
is $1.26 \pm 0.10\%$ with a position angle of $114^{\circ}.2\pm2^{\circ}.2$.
Another set of broad band polarimetry was obtained on Oct. 5.175
and the resulting polarization was found to be $1.32 \pm 0.28\%$ with
position angle $125^{\circ}.2\pm 1$ (Rol, et al. 2002). About 3 days
after these polarization
measurement, ESO VLT-UT3 obtained yet another measurement and
found the \grb\ to be polarized at $0.67\pm 0.23\%$ with position
angle $89^{\circ}.0\pm1^{\circ}.0$ (Covino et al. 2002d). These observations
show marginal evolution of the degree of polarization.

Out spectropolarimetry data was obtained on Oct 5.202 - 5.269, soon after
the observations of Rol et al (2002). The continuum polarization in the
$V$-band synthesized from these observations is $1.67\pm0.10\%$ position
angle of $116.3^{\circ}\pm2^{\circ}$.  
The errors of the position angle of our 
measurement are obtained directly from the RMS scatter of the
polarization position angle at different wavelengths. Despite the fact that
the polarization position angle is consistent with those reported
in Covino et al (2002c, 2002d) and Rol (2002), the degree of polarization
is considerably higher than the corresponding reported values. This suggest
a rapid evolution of the degree of polarization in the early phases. It is
interesting to note that spectropolarimetry and the broad band
polarimetry reported in Covino et al (2002c) and Rol (2002) were obtained
right at the tail of a big bump in the light curve of the \grb\
(e.g. Schaefer et al., 2002). The time evolution of the $V$-band
polarimetry is shown in Figure 2.

\begin{figure}
\figurenum{2}
\epsscale{0.8}
\plotone{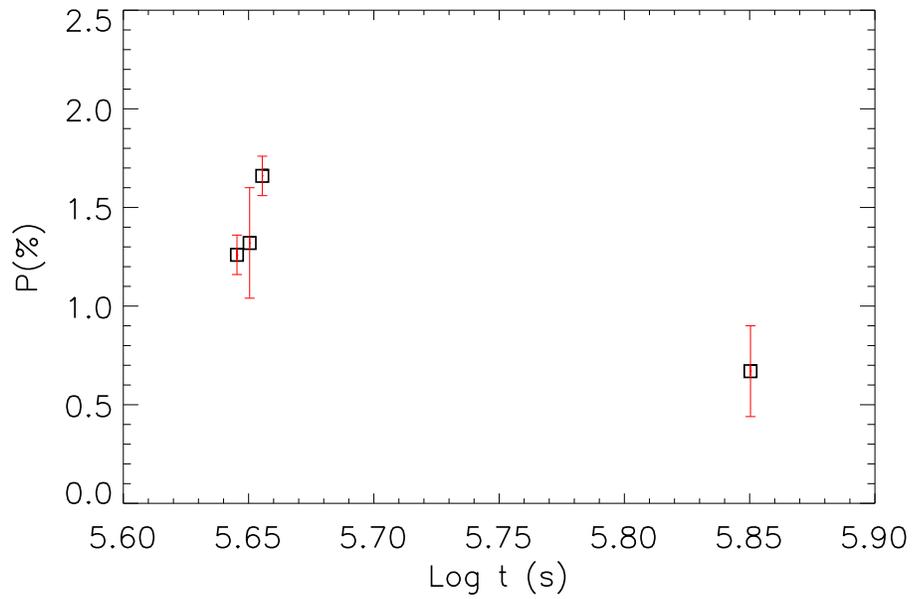} 
\caption{The time evolution of the degree of the V-band polarization. The
first and last data points are from Covino et al (2002c, d), the second
data point is from Rol et al. (2002) and the third data point is from the
current observations. The evolution of the degree of polarization
implies that the detected polarization is at least partially intrinsic
to  GRB 021004.
}
\end{figure}

\subsection{The Polarization across the Strongest Narrow Absorption Lines}

The absorption lines are sharp and strong, and the noise at the line
dips becomes much larger than neighboring wavelength regions.
The instrumental resolution also makes correlations between neighboring
pixels significant in the statistical analysis of the polarization data.
The data are thus re-sampled to bins comparable to the spectral
resolution which is 12\AA. The re-sampling is done using the
weighted integration methods described in Wang et al. (1996) which
takes into account the photon statistical errors
due to each pixel. The binned data are shown in
Figure 3a for the \La\ lines and in Figure 3b for the Si IV and C IV lines.
An apparent increase of the degree of polarization is
observed at the 2$\sigma$ level across the narrow \La\ line at 401.1 nm,
as shown in Figure 3. A similar change is also seen across
the C IV 515.2 nm line, but with even lower significance.
No such sharp change is observed for the stronger \La\ line at 404.0 nm,
or any other strong lines.

\begin{figure}
\figurenum{3}
\epsscale{1.0}
\plottwo{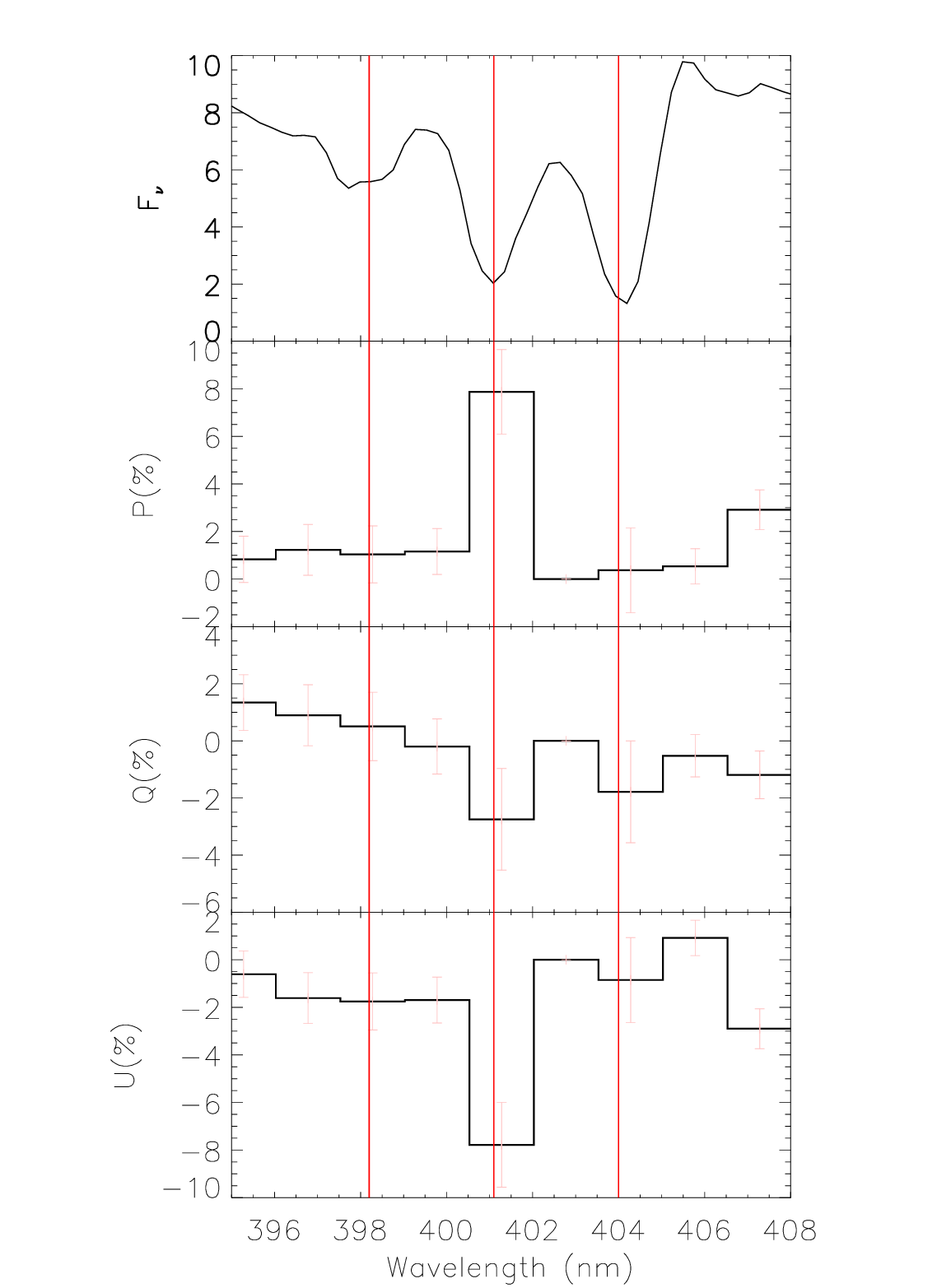}{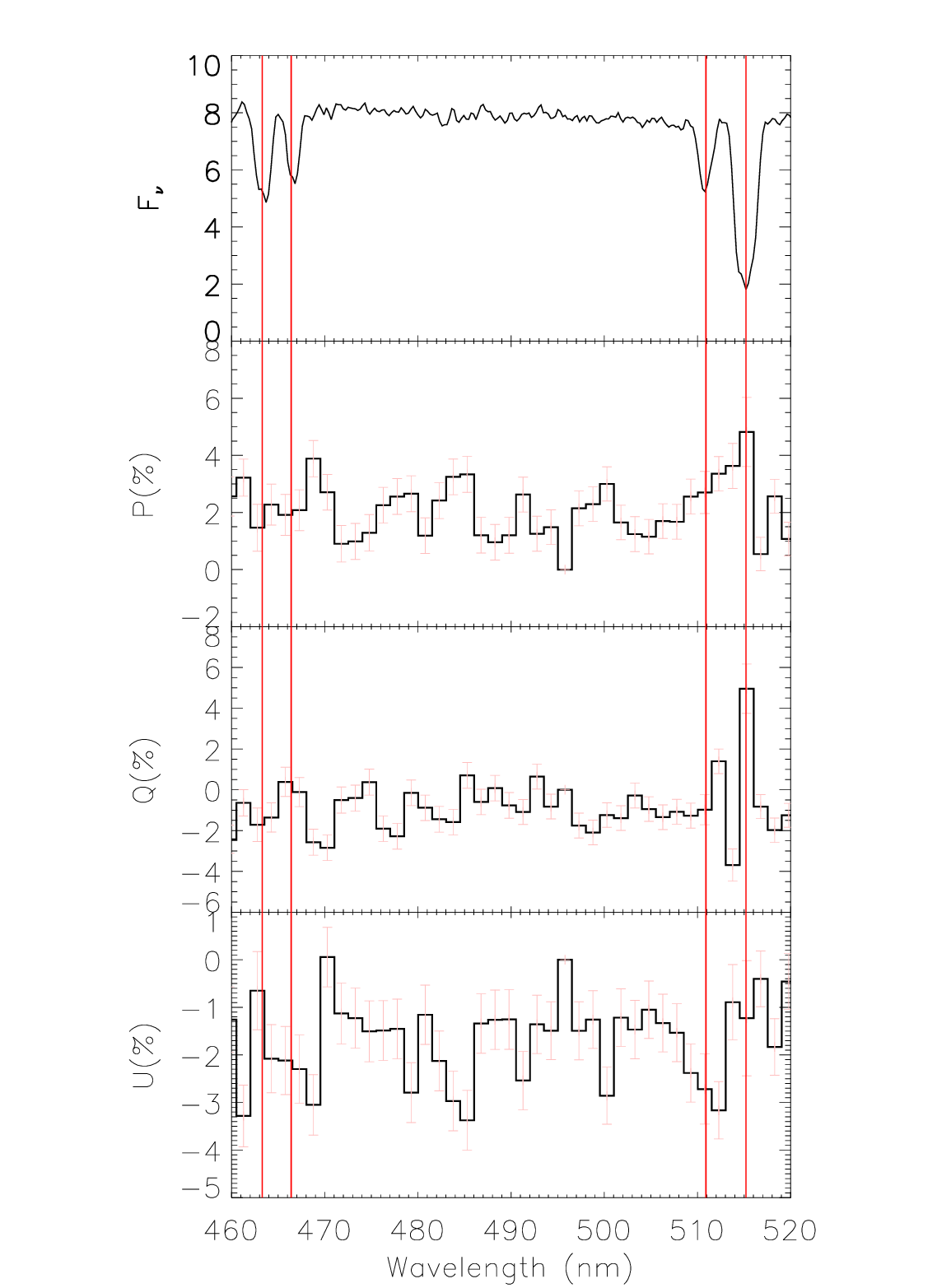}
\caption{Polarization around the strongest narrow lines of \La\ (Left) and
C IV and Si IV (right).
}
\end{figure}

\subsection{The Polarization Blueward of \La 4040A}

We show in Figure 4 the polarization blueward of the
\La\ absorption line at 404.0 nm. A gradual increase of the mean
degree of  polarization toward the blue end of the spectrum is observed.
Despite the loss of sensitivity of the instrument at the very blue end,
polarization as high as 5\% is detected at levels above 2$\sigma$.

In Figure 5, we show the \La\ absorption lines after subtracting a
power-law fit to longer wavelengths.  With the high signal to noise
of these data, essentially all the features in the total flux
spectrum are real.  If we interpret all these features as \La,
it can be seen that the spectrum consists of many
narrow absorption features $\sim$ 1 nm wide to a velocity as high as 40,000
\kms\ from the rest frame defined at z=2.326. Typical separation of the
lines is around  2000-3000 \kms .  Two especially strong absorptions are 
seen at 362 and 368 nm that we interpret as \La\ at
27,000 \kms\ and 31,000 \kms.
In principle, these features could be attributed to a large number of
intergalacitic clouds forming the \La\ forest, or to high velocity 
clumps/shells in the direct neighborhood of the \grb. 

The argument in favor of an intimate connection of the multiple
L$\alpha$ absorption features to the \grb\ comes from the polarization.
Since the increase in the polarization shown in Figure 4
sets in right at the rest 
frame \La , it is reasonable to assume that \La\ is directly 
responsible for the change in polarization (see \S 4.2).
 This supposition finds further support in the polarization change over \La\
in the three strongest components at $\sim$ 3000, 27,000 and 30,000 \kms.
Since intergalactic \La\ absorption systems are incapable of
producing polarization, it thus seems likely that the absorptions blueward
of \La\ are associated with material of corresponding velocity:
material/clumps surrounding the
\grb\ with velocity far in excess of that to be expected from stellar winds.

Figure 5 also shows portion of the spectrum from about 450 to 520 nm 
that contains the strong absorption lines of \ion{C}{4} and \ion{Si}{4}  
with the velocity rest frame chosen to be that of the rest frame of 
\ion{C}{4} 154.9 nm.  Note that the two strongest components of \ion{C}{4}
nearly line up with those of \La.  Also note that there no 
strong high velocity (20,000 -- 30,000 \kms) components to
the \ion{C}{4} and \ion{Si}{4} lines are observed.  This suggests
that the broad decrease in flux and rise in polarization  blueward
of \La\ are due to hydrogen.

\begin{figure}
\figurenum{4}
\epsscale{0.8}
\plotone{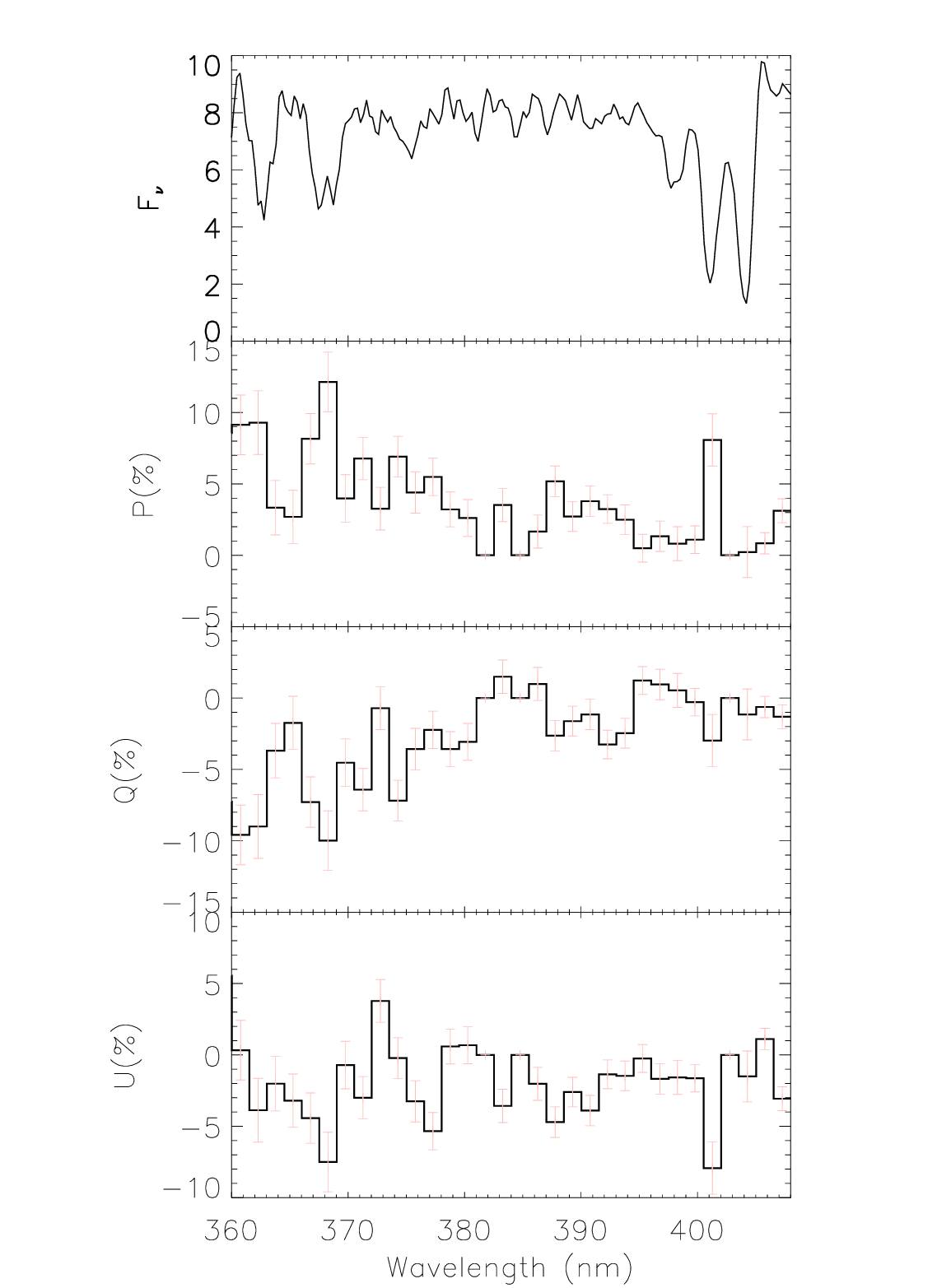}
\caption{Polarization blueward of the \La\ line at 404.0 nm. A gradual
increase of the degree of polarization is observed toward the blue end of
the spectrum (second panel). The degree of polarization is significantly 
higher than that measured in the $V$-band. This suggests that the 
absorption features blueward of the rest frame \La\ (top panel) are 
not entirely due to intervening \La\ intergalactic absorption systems.
}
\end{figure}
\begin{figure}
\figurenum{5}
\epsscale{0.8}
\plotone{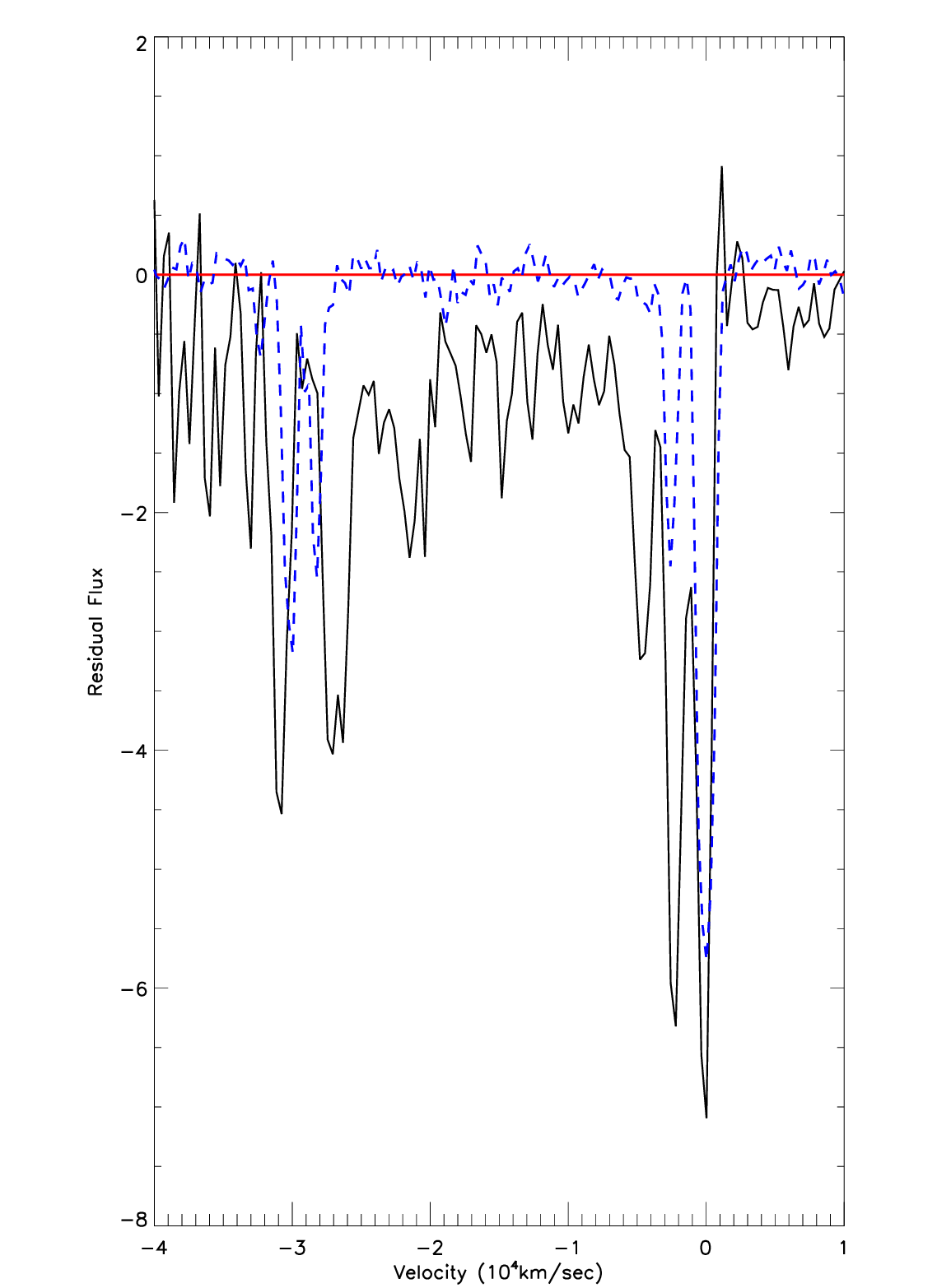}
\caption{The absorption features blue ward of the \La\ line at 404.0 nm
(solid line) plotted in the rest frame of the \La\ absorption line. 
A power law fit to the continuum is subtracted to reveal the absorption 
lines. Also shown is the spectrum showing the \ion{C}{4} and \ion{Si}{4} 
lines from about 450 to 520 nm in the rest frame of \ion{C}{4} 154.9 nm 
(dashed line), where no strong high velocity (20,000 -- 30,000 \kms) 
components are observed.}
\end{figure}

\section{Discussions and Conclusions}

\subsection{Implications for Published Models for $\gamma$-Ray Burst
Polarization}

The optical emission from a \grb\ is produced by synchrotron emission
which is usually highly polarized. The degree of the polarization
depends on the scale of the fluctuations of the magnetic
field in the \grb\ source. When the spatial scales of the fluctuations are
about $N$ times
smaller than the extent of the \grb\ source, the degree of polarization is
reduced by $N$ times
that of the polarization of each single magnetic patch. Gruzinov \&  Waxman
(1999) studied this
case, and found that the integrated  degree of polarization can be around
$P\ \sim\ 60\%/N$. 
The continuum and the $V$-band polarization was observed to be around 1\% for
GRB 021004. It thus
seems that $N\sim 60$ and the typical  scale of the magnetic field is around
1.6\% of that of the
\grb\  source.  This means there are about 3,600 patches on the surface
of the \grb\ optical afterglow.

In the case of the collimated fireballs of Ghisellini \& Lazzati (1999)
and Sari (1999), the polarization has a more complex behavior related
to the break time of the total flux light curve. The polarization is
zero at jet break out. GRB 021004 is obviously associated with very
complicated circum-burst matter where several bumps are observed in the
broad band
light curves. The polarization measurements reported here were obtained
at a time corresponding to the end of a bump of the broad band light
curve. It is not clear if this is merely a coincidence, or of more
fundamental significance.

\subsection{Why Would Any Absorption Features be Polarized ?}

Questions then arise as to why the broad absorption feature blueward of
\La\ is
polarized, and why the polarization varies across
any narrow absorption features? No such variations are expected
within the context of published models for \grb\ polarization.

We propose a very simple model for the observed polarization 
variation across spectral lines. For illustrative purposes, we start
with the patchy models described above and as in Ghisellini \& Lazzati
(1999). We assume the surface of the \grb\ optical afterglow consists
of $N^2$ patches each polarized at around $60\%$. The unabsorbed wavelength
regions will then be polarized at around $60\%/N$.
The absorbing circum-burst matter may not be uniform, but could be
patchy clouds with varying opacity. The patchy structure of the
clouds is expected to absorb different parts of the surface
of the optical afterglow differently. Assuming there are $M^2$
absorption patches
covering the \grb\, and for simplicity that the patches are completely
opaque, the degree of polarization from such a configuration
would be of the order $60\%/\sqrt{(N^2-M^2)}$ for $N \ge M$.
For $N\ <<\ M$, the patchiness of the surrounding medium would have little
effect on the polarization of the \grb.
We see from these equations that $M$ should be of the same order
of $N$ for it to have a significant effect and to account for the observed
$5-10\%$ polarization in the \La\ lines.

It is likely that the patchiness of the circum-burst matter is related to
the \grb\ event itself. The intense high energy photons from the \grb\
ionize the material ahead of the blast wave. Both inhomogeneities of the
GRB shock and the density fluctuations in the circum-burst matter would 
affect the opacity patterns. The patchiness of the
emitting surface of the afterglow shock implies that the opacity of the clouds
reflects the emission patterns of the \grb, but from an epoch earlier
than when the optical afterglow becomes prominent. The difference is
caused by the fact that the ionization starts immediately after the
\grb\ event, which is earlier than the actual
epoch when these observations were made, and the absorption is produced
after the denser part of the circum-burst material recombines to form neutral
hydrogen.

 This brings us to the question of the nature of \grbs. To form blue
absorption lines, the absorbing clumps must be outside the region that 
produces the afterglow. Expansion velocities of the order of 30,000 \kms\ 
far exceed those expected from winds of massive stars or WR stars.
Either the matter is accelerated by the \grb\ or the material 
was accelerated by a stellar explosion (there could be some radiative
acceleration in the latter case). According to Schaefer et al. (2002), 
radiative acceleration to such high velocities would require a location 
of the clouds with the highest velocities at $\sim 10^{16}$ cm.
Even if a wind could be radiatively accelerated to a velocity of 
30,000 \kms\, a Wolf-Rayet star is not a likely candidate.  
If our interpretation of the absorption features blueward of the
rest frame \La\ as high velocity \La\ is correct, then 
the high-velocity hydrogen lines are representative 
of the explosion of a massive star -- a Type II supernova, 
with a hydrogen-rich envelope or wind.

The nature of the high-velocity polarized matter blueward of \La\  
can be further constrained by the afterglow light curves.
Assuming, as found by Schaefer et al. (2002) that the after glow is produced
at distances between $10^{16}$ to 10$^{17}$ cm about 1 day after the burst.
The particle densities of the surrounding medium are of the order of 
30 to 100 cm$^{-3}$. For those densities, the \grb\ would fully ionize the 
matter and, using standard recombination theory, 
hardly any \ion{H}{1}, \ion{C}{4}, or \ion{Si}{4} lines would be observable.
Note that the standard recombination theory may be not be appropriate
in this context because the electron gas is highly non-thermal, thus
violating a basic assumption for calculating the recombination cross sections.
These caveats aside, it is thus difficult 
to see how a circum-burst medium consisting only of a stellar wind
could account for both the high-velocity, high-ionization lines
and for the even higher velocity hydrogen suggested by the polarization.
Rather, it seems that there must be some rather high-density matter
in the vicinity of the burst that would allow rapid recombination
of some high-velocity hydrogen into the ground state.

The overall smooth variation of the \La\ polarization 
blueward of the rest frame suggests that the opacity of the absorbing 
clumps is affected by the \grb, and that the material responsible for 
the absorption is at least partially due to matter in the vicinity of 
the \grb. In contrast to the density expected for a stellar wind at distances
for which some ions could survive the burst, a supernova envelope
could provide both the very high velocities deduced for some of the 
\La\ and sufficiently high densities that recombination could be rapid.  
For supernova envelopes, we expect densities of about 
$10^{8}$ to 10$^{9}$ particles cm$^{-3}$ at a distance of $10^{16}$ cm 
resulting in very small recombination times.  

The scenario that emerges to account for the polarization is
the presence of the ejecta of a Type II supernova with a 
thin layer of hydrogen ejected to velocities as high as 40,000 \kms. 
Such high velocity material is not unprecedented, but is indeed observed 
in some Type II supernovae. In order for the supernova material that
causes the high-velocity \La\ absorption and the polarization
blueward of the rest frame \La, the supernova material must lie outside
the afterglow shock when the observations were made about 1 day after
the \grb.  To reach a distance in excess of $10^{16}$ cm, the supernova 
explosion must have preceded the \grb\ by several weeks. But note again
that this conclusion is based on the assumption that the after glow was
produced at a distance about $10^{16}$ cm from the the center of the
GRB.

If this picture is correct, the resulting densities at $>\ 10^{16}$ cm are
of order $10^{8}$ cm$^{-3}$, i.e. in contradiction to the densities
derived from the light curve. This may suggest a medium that has both low and
high densities at about the same distance. This and the intrinsic line widths 
of the high-velocity \La\ components (1000 \kms) that are small
compared  to the overall velocity (30,000 \kms) indicate a very 
clumpy medium.  
Alternatively, high velocity bullets penetrating the ISM prior to the
classical supernova shell may also be possible.

Our polarization observations have thus led to the suggestion that
a supernova preceded GRB 021004 by weeks to months.  This is
very reminiscent of the ``supernova" hypothesis of Vietri \& Stella
(1998, 1999) in which a neutron star with mass above the non-rotating
limit eventually loses its rotational support and collapses to
make a black hole.  This model has been invoked to account for
transient X-ray Fe lines in GRB~991216 by Vietri et al. (2001;  see also
Lazzati, Campana \& Ghisellini 1999), B{\"o}ttcher \& Fryer(2001),
and B{\"o}ttcher, Fryer, \& Dermer(2002),
and in the context of pulsar wind models for \grbs\ by
K{\" o}nigl \& Granot(2002) and Wang, Dai \& Lu (2002).
One drawback of this model is the uncertainty in the initial
rotation state and hence the plausibility of the initial ``supermassive"
neutron star rotation state and the uncertainty of the resulting
time to collapse.  Nevertheless, our interpretation of the behavior of
the flux blueward of the rest frame \La\ and its polarization in GRB~021004
has led us to a similar conclusion, that a hydrogen-rich supernova
exploded several weeks to months prior to the \grb..   

Our spectropolarimetry has raised the intriguing possibility that
GRB 021004 is associated with a supernova. Whether or not this picture
ultimately prevails, these observations clearly show that prompt  
spectropolarimetry is a powerful tool in the quest to resolve
the mystery of the origin of \grbs.

{}

\end{document}